\documentclass[aps,pra,twocolumn,showpacs,longbibliography]{revtex4-2}
\usepackage{graphicx} 
\usepackage{amsmath}
\usepackage{graphicx,epstopdf}
\usepackage{tabu}
\usepackage{multibib}
\usepackage{gensymb}
\newcommand{\be}{\begin{equation}}
\newcommand{\ee}{\end{equation}}
\newcommand{\bea}{\begin{eqnarray}}
\newcommand{\eea}{\end{eqnarray}}
\newcommand{\bse}{\begin{subequations}}
\newcommand{\ese}{\end{subequations}}

\usepackage{color}
\usepackage[colorlinks,bookmarks=false,citecolor=darkblue,linkcolor=red,urlcolor=blue]{hyperref}

\definecolor{darkred}{rgb}{0.7,0.0,0.0}

\definecolor{darkblue}{rgb}{0,0.02,0.45}

\definecolor{darkgreen}{rgb}{0.02,0.45,0.0}

\definecolor{violet}{rgb}{0.8,0.2,0.6}

\begin{document}
\title{Negative thermal expansion and itinerant ferromagnetism in Mn$_{1.4}$Fe$_{3.6}$Si$_{3}$}
\author{Vikram Singh}
\email{vikram51128@gmail.com}
\affiliation{School of Physics, Indian Institute of Science Education and Research Thiruvananthapuram-695551, India}
\author{R. Nath}
\email{rnath@iisertvm.ac.in}
\affiliation{School of Physics, Indian Institute of Science Education and Research Thiruvananthapuram-695551, India}
\date{\today}

\begin{abstract} We report the thermal expansion, critical behavior, magnetocaloric effect (MCE), and magnetoresistance ($MR$) on the polycrystalline Mn$_{1.4}$Fe$_{3.6}$Si$_{3}$ compound around the ferromagnetic transition. 
A large negative volume thermal expansion ($\alpha_{\rm V}\sim -20 \times 10^{-6}$~K$^{-1}$) is observed across the transition temperature with a strong anisotropic variation of lattice parameters in the $ab$-plane. The anisotropic magnetoelasticity arises from the competition between magnetic ordering and structural deformation which could be responsible for the large MCE ($\Delta S_{\rm m} \simeq -6$~J/Kg-K) across the magnetic transition in this compound.
The large and negative $MR$ ($\sim -3\%$ in 80~kOe) is also observed at the transition temperature which can be attributed to the suppression of spin disorder. Further, the Rhodes-Wolfarth ratio (RWR~$> 1$) and identical field dependence of $MR$ and MCE isotherms indicate the itinerant character of the $3d$ electrons. The critical exponents
determined from the analysis of magnetization and MCE are consistent with the quasi-two-dimensional (2D) Ising model with long range exchange interactions which decays as $J(r)\sim r^{-3.41}$. This unconventional quasi-2D Ising character with long-range interactions can be ascribed to strong $ab$-plane anisotropy and the delocalized $3d$ electrons in the studied compound.
\end{abstract}
\maketitle

\section{Introduction}
In the past decades, research on materials with almost zero thermal expansion (ZTE) and negative thermal expansion (NTE) has been significantly enhan ced due to their potential technological applications~\cite{Takenaka013001}.
Recently, the large or giant NTE has been found in a variety of materials including oxides, intermetallics, alloys, antipervoskites, fluorides, and organometallic frameworks 
over a wide range of temperature~\cite{Mary90,Guillaume235,vanSchilfgaarde46,Takenaka131904,Song4690,Chen11114,Huang11469, Goodwin794,Song85,Hu1375836,Greve15496,Chen3522}. Among these materials, the NTE in compounds with large magnetocaloric effect (MCE) has attracted special attention due to strong magnetoelastic coupling. In these materials, the NTE occurs as a result of the volume expansion accompanied by magnetic transition, known as magnetovolume effect (MVE) which dominates over the conventional phononic thermal expansion~\cite{Takenaka013001}. Here, a disordered magnetic phase with smaller volume transforms to a ordered magnetic structure with larger volume as the temperature decreases, causing a negative thermal expansion~\cite{vanSchilfgaarde46}. Prominent examples include La(Fe,Si)$_{13}$, Tb(Co,Fe)$_{2}$, Mn$_{3}$Ge, $R_{2}$Fe$_{17}$ ($R$ = rare earth elements), (Hf,Ta)Fe$_{2}$, (Sc,Ti)Fe$_{2}$, MnCoGe etc~\cite{Huang11469,Song602,Song6236,Pablo184411,Li224405,Song275,Ren17531}. Furthermore, the manipulation or control of NTE via different routes such as chemical substitution, nano-crystallization, introduction of disorder, and application of external pressure and magnetic field provides extra working flexibility and advantage in multifunctional applications in these materials.

Basically, these MCE materials are divided into two categories based on the nature of the magnetic transition: first order transition and second order transition. In the first order, the magnetic transition is accompanied with an abrupt structural deformation which gives rise to the sharp and large MVE at the transition. The colossal NTE in such materials is mainly associated with the abrupt enhancement of unit cell volume at the transition and is being investigated extensively in the recent days~\cite{Hu1375836}. On the other hand, the pronounced NTE across a continuous second order transition has been observed only in few selected MCE materials including the Invar alloys and intermetallic compounds La(Fe,Co,Si)$_{13}$ due to the itinerant ferromagnetism and magnetoelastic coupling~\cite{Guillaume235,Huang11469,Takenaka013001}.

MnFe$_{4}$Si$_{3}$ undergoes a paramagnetic (PM) to ferromagnetic (FM) transition upon cooling near room temperature without altering the crystal symmetry (hexagonal) and shows large MCE around the transition. During cooling, the crystal lattice experiences an expansion along the $a$-axis, while the unit cell volume shows only a weak plateau across the transition temperature~\cite{Hering7128}. Subsequently, Herlitschke et al reported that MnFe$_{4}$Si$_{3}$ exhibits magnetoelasticity across the magnetic transition~\cite{Herlitschke094304}. Similar to the other systems, one can tune/control the strength of magnetoelastic coupling in the compound under investigation by chemical substitution~\cite{Franco414004,Fujita104416,Dung092511,Dewei224105}. Therefore, it would be interesting to investigate the thermal expansion behaviour of the Mn$_{1+x}$Fe$_{4-x}$Si$_{3}$ series along with the magnetic properties across the PM-FM transition.
Recently, we have investigated the Mn$_{1+x}$Fe$_{4-x}$Si$_{3}$ series for $x = 0$ to $1$ and found that the magnetic transition can be tuned continuously from above room temperature to lower temperatures with increasing $x$. A detailed analysis of magnetization and MCE for two compositions ($x= 0.0$ and 0.2) suggest that this series of compounds can be used for continuous room temperature magnetic refrigeration purpose. The critical analysis of the magnetization and MCE across the PM-FM transition for $x =0$ and $0.2$ yield similar but unconventional critical behavior, the origin of which is not yet clear~\cite{Singh6981}. Further, as there is a magnetoelastic coupling, one expects a correlation between thermal expansion and magnetic behaviour which was overlooked in our previous study. 

In the present work, we choose another composition $x =0.4$ (Mn$_{1.4}$Fe$_{3.6}$Si$_3$) and thoroughly investigated the thermal expansion, magnetization, magnetoresistance ($MR$), and critical behavior around $T_{\rm C} \simeq 254$~K. Our experiments demonstrate that Mn$_{1.4}$Fe$_{3.6}$Si$_3$ exhibits a large NTE across the magnetic transition and almost ZTE below the transition. We observed that the critical behavior is not altered with varying $x$ and the values of critical exponents are found to be similar to other itinerant ferromagnets reported in the literature. This unconventional critical behavior is explained in terms of strong anisotropy and itinerant character of the $3d$ electrons in the compound.

\section{Experimental Details}
Polycrystalline sample of Mn$_{1.4}$Fe$_{3.6}$Si$_{3}$ is synthesized by arc melting followed by thermal annealing in vacuum at 950~$^{o}$C for five days. To check the phase purity, powder x-ray diffraction (XRD) is performed using PANalytical X’Pert Pro diffractometer with Cu K$_{\alpha}$-source ($\lambda = 1.5406$~\AA). To analyze the structural changes across the magnetic transition, temperature dependent powder XRD measurements are performed in a temperature range 300~K to 15~K. For this purpose, an Oxford PheniX closed cycle helium cryostat is used as an attachment to the diffractometer. Rietveld refinement of the XRD data is performed using FullProf software package~\cite{Rodriguez55}. The $dc$- and $ac$- magnetic measurements as a function of temperature and magnetic field are performed using Vibrating Sample Magnetometer (VSM) and $ac$-susceptibility options of 9~Tesla PPMS (Quantum Design), respectively. While measuring magnetic isotherms at and below $T_{\rm C}$, the demagnetization field has been subtracted from the applied field following the procedure described in Ref.~[\onlinecite{Kaul1114}]. The temperature dependent (4-300~K) resistivity is measured using four probe method in a home made resistivity set-up attached to a cryostat (M/s. OXFORD Instrument, UK) with 8~Tesla superconducting magnet. For the magnetoresistance measurement, the magnetic field is applied in the longitudinal geometry. 
\section{Results and Discussion}
\begin{figure}[t]
\begin{center}
\includegraphics[width=\linewidth]{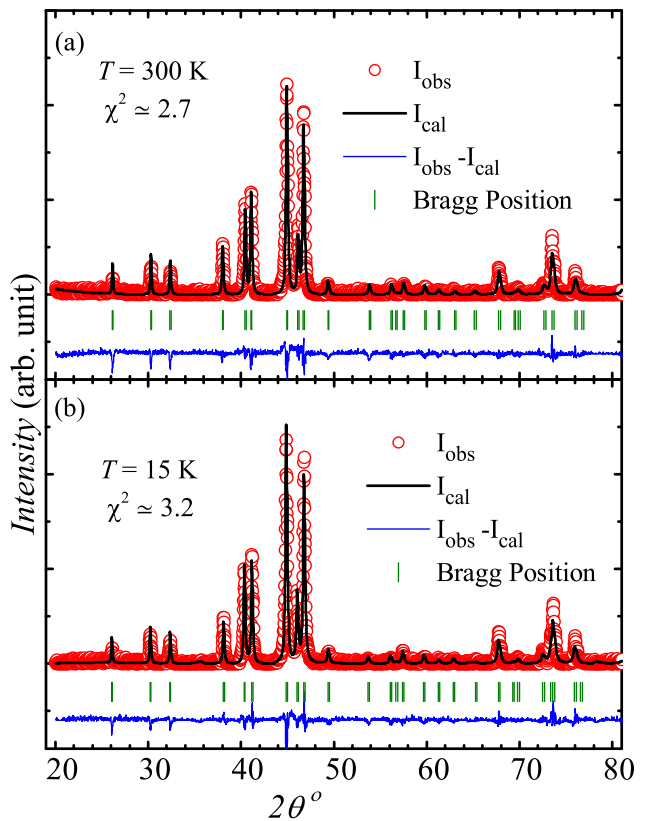}
\end{center}
\caption{\label{Fig1} The powder XRD patterns of Mn$_{1.4}$Fe$_{3.6}$Si$_{3}$ (a) at room temperature and (b) at $T = 15$~K. The solid black line represents the Rietveld refinement of the experimental data, the green vertical bars correspond to Bragg positions, and the bottom blue line represents the difference between observed and calculated intensities.}
\end{figure}
\begin{figure}[t]
\begin{center}
\includegraphics[width=\linewidth]{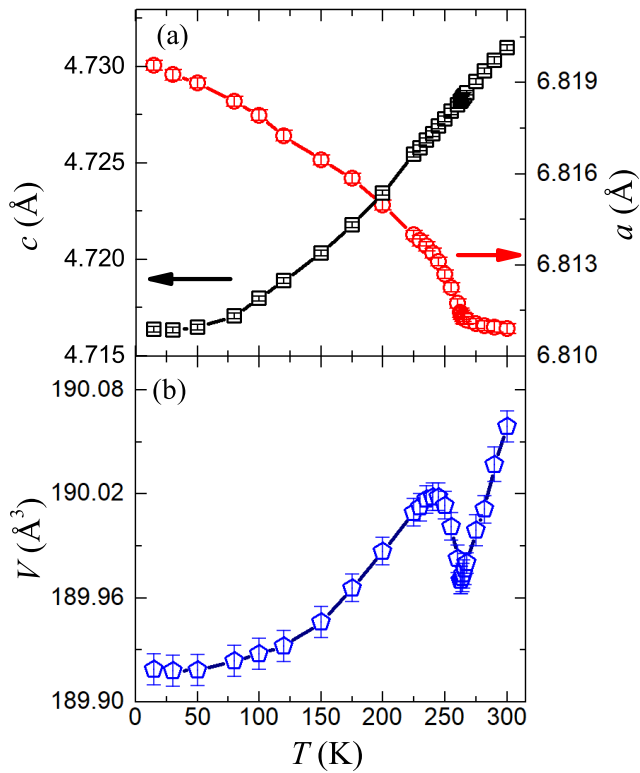}
\end{center}
\caption{\label{Fig2} The variation of lattice parameters (a) $a$ and $c$ and (b) unit cell volume $V$ with temperature, obtained from the Rietveld refinement.}
\end{figure}
\begin{figure}[t]
\begin{center}
\includegraphics[width=\linewidth]{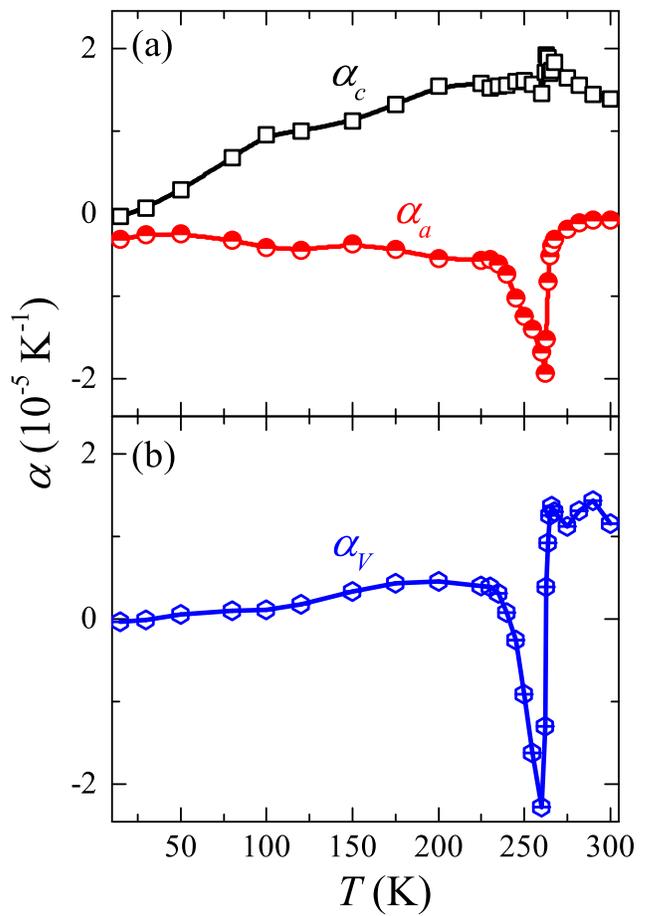}
\end{center}
\caption{\label{Fig3} The variation of thermal expansion co-efficients with temperature for (a) $a$ and $c$ and (b) $V$.}
\end{figure}
\subsection{X-ray Diffraction}
Figure~\ref{Fig1}(a) and (b) present the Rietveld refinement of the powder XRD patterns of Mn$_{1.4}$Fe$_{3.6}$Si$_{3}$ measured at $T = 300$~K and 15~K, respectively. Both the patterns can be refined with space group $P6_{3}/mcm$ of the hexagonal crystal symmetry, without detecting any extra peak. The lattice parameters [$a = 6.8112(4)$~\AA, $c = 4.7351(3)$~\AA, and unit cell volume $V = 190.06(2)$~\AA$^{3}$] obtained from the refinement of the room temperature XRD are in good agreement with our previous report~\cite{Singh6981}.

The variation of lattice parameters ($a$, $c$, and $V$) as a function of temperature is shown in Fig.~\ref{Fig2}. As the temperature is lowered from 300~K, $a$ remains almost constant down to $\sim 270$~K and then exhibits a continuous rise below 270~K, the temperature below which the PM to FM transition sets in. In contrast, $c$ decreases monotonically from 300~K without showing any clear anomaly at the transition temperature. The overall unit cell volume features a sharp dip at the transition temperature. This indicates a strong coupling of lattice deformation in the $ab$-plane with the magnetization or magnetic transition.

Further, to explore the relation between the lattice expansion and magnetism, we have extracted the thermal expansion co-efficients $\alpha_{\rm A} = \frac{1}{A}\left(\frac{\partial A}{\partial T}\right)_{\rm P}$, where $A$ stands for the lattice parameters ($a$, $c$, and $V$) and $P$ is the pressure. The obtained thermal expansion co-efficients are displayed in Fig.~\ref{Fig3} as a function of temperature. It is observed that at the transition temperature, $\alpha_a$ exhibits a large and negative thermal expansion (NTE), whereas $\alpha_c$ shows only a weak and positive anomaly. These results indicate that the transition is accompanied by an anisotropic variation of the unit cell or lattice distortion where it expands rapidly in the $ab$-plane and contracts weakly along the $c$-axis. This leads to an overall NTE of the unit cell volume in a wide temperature range 230~K - 265~K. It reaches a minimum value of $\alpha_{\rm V}\simeq -20\times10^{-6}$~K$^{-1}$ at the transition temperature, which is comparable to the value reported for large NTE materials such as manganese antiperovskites, itinerant La(Fe,Si)$_{13}$ compounds etc~\cite{Song4690,Takenaka131904,Huang11469}. As the width of the NTE peak decides the practical use of the compound, one can widen the peak width and tune the transition temperature by introducing disorder in the system through quenching or reduction of particle size and by chemical substitutions~\cite{Song4690}. Below the transition, $\alpha_V$ has an almost zero (ZTE) or very small value which remains almost constant ($\sim 2 \times 10^{-6}$~K$^{-1}$) with temperature, similar to Invar alloys \cite{Guillaume235,Wasserman1990} due to the opposite variation of lattice parameters $a$ and $c$ [see Fig.~\ref{Fig2}(a)]. The NTE could be related to softening of the phonon modes across the magnetic transition~\cite{Herlitschke094304,Gruner057202}. Moreover, the anisotropic thermal expansion in Mn$_{1.4}$Fe$_{3.6}$Si$_3$ across the transition is quite consistent with strong magnetocrystalline anisotropy reported in the parent compound MnFe$_{4}$Si$_{3}$, where the easy axis lies in the $ab$-plane~\cite{Hering7128,Biniskos104407}. This indicates that the magnetic transition involves a strong correlation between magnetic and structural degrees of freedom.

For a second-order transition, the change in volumetric thermal expansion coefficient ($\Delta \alpha_V$) at the transition is directly related to the pressure dependence of $T_{\rm C}$ through the Ehrenfest relation 
\begin{equation}
\frac{dT_{\rm C}}{dP}= \frac{\Delta \alpha_{\rm V}~V_{\rm mol}~T_{\rm C}}{\Delta C_{\rm P}},~{\rm near}~P = 0.
\label{eq.1} 
\end{equation}
Here, $\Delta C_{P}$ is the change in heat capacity at constant pressure at $T_{\rm C}$ and $V_{\rm mol}$ is the molar volume. From Eq.~\eqref{eq.1}, one can estimate the variation of $T_{\rm C}$ with pressure for known values of $\Delta C_{P}$ and $\Delta \alpha_V$. Unfortunately, we don't have $C_{\rm P}$ data of our compound for doing such an analysis. Nevertheless, the observed negative value of $\Delta \alpha_V$ implies negative $\frac{dT_{C}}{dP}$, where the transition shifts to lower temperatures with increasing external pressure. Interestingly, this finding is consistent with the pressure dependent study on MnFe$_{4}$Si$_{3}$, where $\frac{dT_{C}}{dP}= -15$~K/GPa is observed~\cite{Eich096118}.

\begin{figure}[t]
\begin{center}
\includegraphics[width=\linewidth]{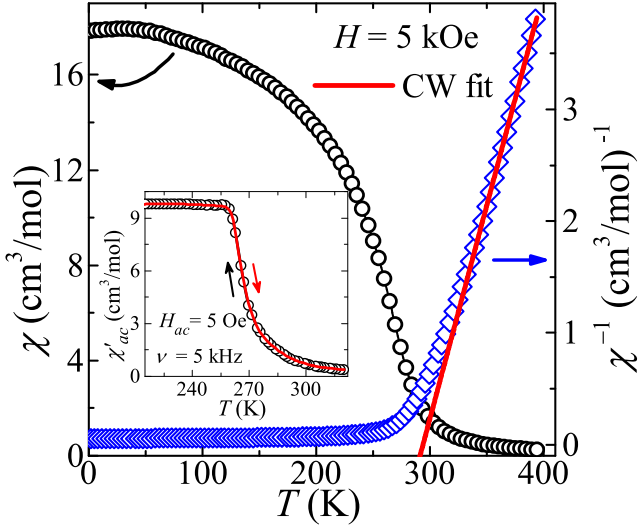}
\end{center}
\caption{\label{Fig4} Temperature dependent magnetic susceptibility $\chi(T)$ and its inverse $\chi^{-1}(T)$ measured in a magnetic field of 5~kOe. Inset: $\chi^{'}_{\rm ac}(T)$ measured during cooling and warming.}
\end{figure}
\begin{figure}[t]
	\begin{center}
		\includegraphics[width=\linewidth]{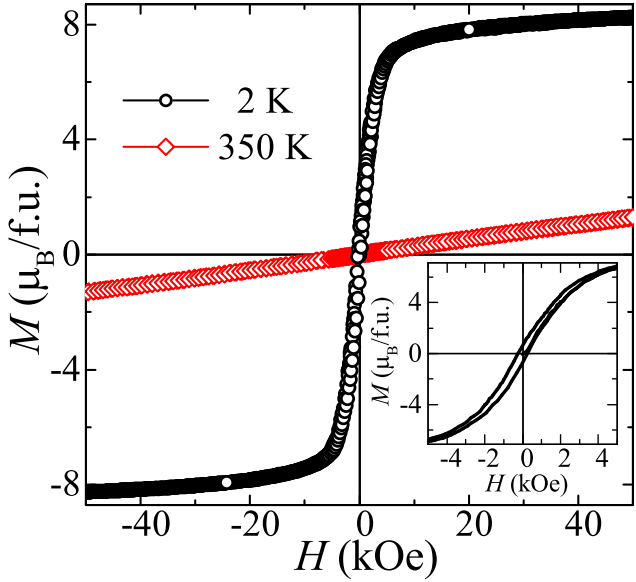}
	\end{center}
	\caption{\label{Fig5} Isothermal magnetization [$M(H)$] measured at $T=350$~K and 2~K. Inset: the $M(H)$ curve $T = 2$~K is magnified in the low field regime to highlight the hysteresis.}
\end{figure}
\subsection{Magnetization}
Temperature dependent $dc$-magnetic susceptibility [$\chi(T)$] and its inverse [$\chi^{-1}(T)$] for Mn$_{1.4}$Fe$_{3.6}$Si$_{3}$ measured in an applied field of $H=5$~kOe are plotted on the left and right $y$-axes, respectively in Fig.~\ref{Fig4}. With decrease in temperature, $\chi(T)$ exhibits a sharp raise below $300$~K corresponding to the onset of PM to FM transition. The real part of $ac$-susceptibility [$\chi_{ac}^{'}(T)$] measured in an $ac$ field of $H_{\rm ac} = 5$~Oe and frequency $\nu=5$~kHz during cooling and warming is shown in the inset of Fig.~\ref{Fig4}. No thermal hysteresis is observed around the transition, indicating second order nature of the transition. To estimate the magnetic parameters, the high temperature part of $\chi^{-1}(T)$ is fitted using Curie-Weiss law, $\chi(T)={C}/({{T-\Theta_{\rm CW}}})$. Here, $C$ is Curie constant and $\Theta_{\rm CW}$ is the paramagnetic Curie temperature. A linear fit ($T\ge 350$~K) yields $\Theta_{\rm CW} \simeq 291.4$~K and effective paramagnetic moment $\mu_{\rm eff} \simeq 2.31\mu_{\rm B}$ per transition metal atom using $C = N_{A}\mu_{\rm B}^{2}\mu_{\rm eff}^{2}/3k_{B}$, where $N_{\rm A}$ is the Avogadro number. These parameters are consistent with the previous report~\cite{Singh6981}.

Figure~\ref{Fig5} shows the isothermal magnetization [$M(H)$] measured at $T=350$~K and 2~K. The straight line behavior of $M(H)$ curve without any saturation at 350~K represents the typical PM state. On the other hand, the $M(H)$ curve at 2~K shows a sharp increase at low fields with a weak hysteresis and a small coercive field of $H_{\rm cor}\simeq250$~Oe, typically expected for a soft ferromagnet. At high fields, magnetization saturates and the value of saturation moment per transition metal atom is estimated to be $\mu_{\rm S}\simeq1.61~\mu_{\rm B}$ from the $y$-intercept of the linear fit to the Arrott plot ($M^{2}$ vs $H/M$) in the high field regime. It is interesting to note that the value of $\mu_{\rm eff}$ is relatively larger than $\mu_{\rm S}$ which is a possible indication of the itinerant character of $3d$ electrons in Mn$_{1.4}$Fe$_{3.6}$Si$_{3}$. Therefore, we calculated the Rhodes-Wolhfarth ratio (RWR~$= q_{\rm c}/q_{\rm s}$) where $q_{\rm c}$ and $q_{\rm s}$ are the number of magnetic carriers per atom deduced from  $\mu_{\rm eff}$ and low temperature saturation moment $\mu_{\rm S}$, respectively~\cite{Rhodes247}. RWR is typically used to distinguish between the localized and itinerant characters of a ferromagnet~\cite{Wohlfarth113,Pramanik214426}. In the case of a localized ferromagnet, RWR should be close to 1, whereas for a itinerant ferromagnet the saturation magnetization is less than the fully polarized moment producing RWR~$>1$. For Mn$_{1.4}$Fe$_{3.6}$Si$_{3}$, we calculated $q_{\rm c} = 1.52$ from $\mu_{\rm eff}$ using the relation $\mu_{\rm eff}^{2} = q_{\rm c}(q_{\rm c} + 2)$ and $q_{\rm s} = 0.89$ from $\mu_{\rm S}$ at 2~K using $\mu_{\rm S}^{2} = q_{\rm s}(q_{\rm s} + 2)$. Thus, the obtained RWR = $1.71>1$ indicates itinerant character of the ferromagnet Mn$_{1.4}$Fe$_{3.6}$Si$_{3}$.

\subsection{Magnetocaloric Effect and Critical Behaviour}
\begin{figure}[hbt]
	\begin{center}
		\includegraphics[width=\linewidth]{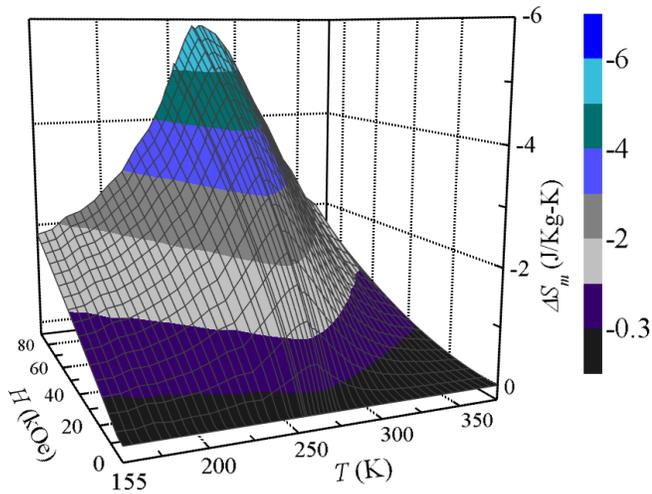}
	\end{center}
	\caption{\label{Fig6} Temperature and magnetic field dependent 3D plots of magnetic entropy change ($\Delta S_{\rm m}$).}	     
\end{figure}
The MCE for the polycrystalline Mn$_{1.4}$Fe$_{3.6}$Si$_3$ is obtained in term of magnetic entropy change ($\Delta S_{\rm m}$). $\Delta S_{\rm m}$ is calculated from the magnetic isotherms measured between 150 - 350~K using the standard expression derived from the Maxwell relation:
 \begin{equation}
	\Delta S_{\rm m}=\int_{H_{i}}^{H_{f}}\dfrac{dM}{dT}dH.
	\label{eq.2}
\end{equation}
The results of our $\Delta S_{\rm m}$ calculation are presented in Fig.~\ref{Fig6} as a function of magnetic field and temperature in the 3D-plot. The $\Delta S_{\rm m}$ curves exhibit a gradual variation with temperature and show a maximum at $T_{\rm C}$. The shape of $\Delta S_{\rm m}$ plots indicate a typical second order transition at $T_{\rm C}$~\cite{Law2680,Islam134433}. The maximum value of $\Delta S_{\rm m} \simeq -6$~J/kg-K is achieved for a field change of 90~kOe. The relative cooling power ($RCP= |\Delta S_{\rm m}^{\rm pk}\times\delta T_{\rm FWHM}|$) is another important parameter for a magnetic refrigeration material and it corresponds to the amount of heat transferred between the source and sink of a refrigerator. Here, $\Delta S_{\rm m}^{\rm pk}$ and $\delta T_{\rm FWHM}$ are the maximum value of $\Delta S_{\rm m}$ at the peak position and the full width at half maximum of the $\Delta S_{\rm m}(T)$ curve, respectively. $\Delta S_{\rm m}^{\rm pk}$ and $\delta T_{\rm FWHM}$ as a function of $H$, obtained from Fig.~\ref{Fig6} are plotted in Fig.~S7(a) of the Supplementary Material. The $RCP$ value estimated for each $\Delta S_{\rm m}(T)$ curve at different magnetic fields is shown in Fig.~S7(b) of the Supplementary Material. It is found to increase with field and reaches a maximum value $\sim 709$~J/Kg for 90~kOe. These values of $\Delta S_{\rm m}$ and $RCP$ are comparable to the values reported for other compositions in Mn$_{1+x}$Fe$_{4-x}$Si$_{3}$ series~\cite{Singh6981}.

\begin{table*}[hbt]
	\caption{\label{arttype} The critical exponents ($\beta$, $\gamma$, $\delta$, and $n$) and $T_{\rm C}$ of Mn$_{1+x}$Fe$_{4-x}$Si$_{3}$ obtained from the modified Arrott plot (MAP), Kouvel-Fisher (KF) plot, critical isotherm (CI), and magnetocaloric effect (MCE)/relative cooling power (RCP) analysis across the PM-FM transition. For comparison, critical exponents corresponding to different theoretical models are also listed.}
	\begin{ruledtabular}	
		\begin{tabular}{ccccccccccc}
			\\ System & Method & $ \beta $  & $ \gamma $ &  $\delta$  & $n$ & &$T_{\rm C}$& Ref. \\ \hline\\ 
			$x = 0.4$&MAP &0.304(3)&1.445(4)&5.75(4)&0.602&&254.0(2)&This work\\
			&KF&0.301(4)&1.441(4)&5.78(7)&0.60&&253.9(1)&  \\
			&CI& --&--&5.71(6)&--&&254&\\
			&MCE/RCP& --&--&5.69(7)&0.601&&254&\\
			\\
			$x = 0.2$&MAP&0.304(3)&1.445(4)&5.75(4)&0.602&&278.17(3)&\cite{Singh6981}\\
			&KF&0.301(1)&1.45(1)&5.77(4)&0.600 &&278.1(1)&  \\
			&CI& --&--& 5.64(3)&--&&278&\\
			&MCE/RCP & --&--&5.73(13)&0.607&&278&\\
			\\
			$x = 0.0$&MAP &0.308(3)&1.448(5)&5.641(4)&0.606&&309.60(2)&\cite{Singh6981}\\
			&KF&0.303(4)&1.451(4)&5.77(7)&0.603&&309.7(1)&\\
			&CI & --&--&5.644(9)&--&&309.6&\\
			&MCE/RCP& --&--&5.70(7)&0.604&&309.6&\\
			\\
			Mean-field model &Theory& 0.5& 1.0& 3.0&&&&\cite{Kaul5}\\
			3D-Heisenberg model&Theory&0.365&1.386&4.80&&&&\cite{Kaul5}\\
			3D-Ising model&Theory&0.325 & 1.241& 4.82&& &&\cite{Kaul5}\\
			3D-XY&Theory&0.345& 1.316& 4.81&& &&\cite{Kaul5}\\
		\end{tabular}
	\end{ruledtabular}
\end{table*}
The critical exponents that reflect the universality class of the spin system are determined using the magnetic isotherms and MCE data following the procedure reported in Ref.~[\onlinecite{Singh6981}]. Please see the Supplementary Material for details of the critical analysis. In Table~I, the critical exponents obtained from different techniques for compositions $x = 0.0$, 0.2, and 0.4 are listed along with the theoretically expected values for different universality classes. Adopting the iterative method with the modified Arrott plot (MAP) and from magnetic isotherm at $T_{\rm C}$, the critical exponents are obtained to be $\beta = 0.304$, $\gamma = 1.4445$, and $\delta \simeq 5.71$ with $T_{\rm C} \simeq 254$~K. These values are further confirmed from the Kouvel-Fisher (KF) method and Widom scaling relation. These values of exponents are identical to the other compositions in the Mn$_{1+x}$Fe$_{4-x}$Si$_{3}$ series and do not fall under any conventional universality class~\cite{Singh6981}. 
Similar unconventional critical behavior has also been observed in various ferromagnets~\cite{Reisser83,Cheng087205,Zhou077206,Bhattacharyya184414,Tateiwa064423,SU1222}.
\begin{figure}[hbt]
	\begin{center}
		\includegraphics[width=\linewidth]{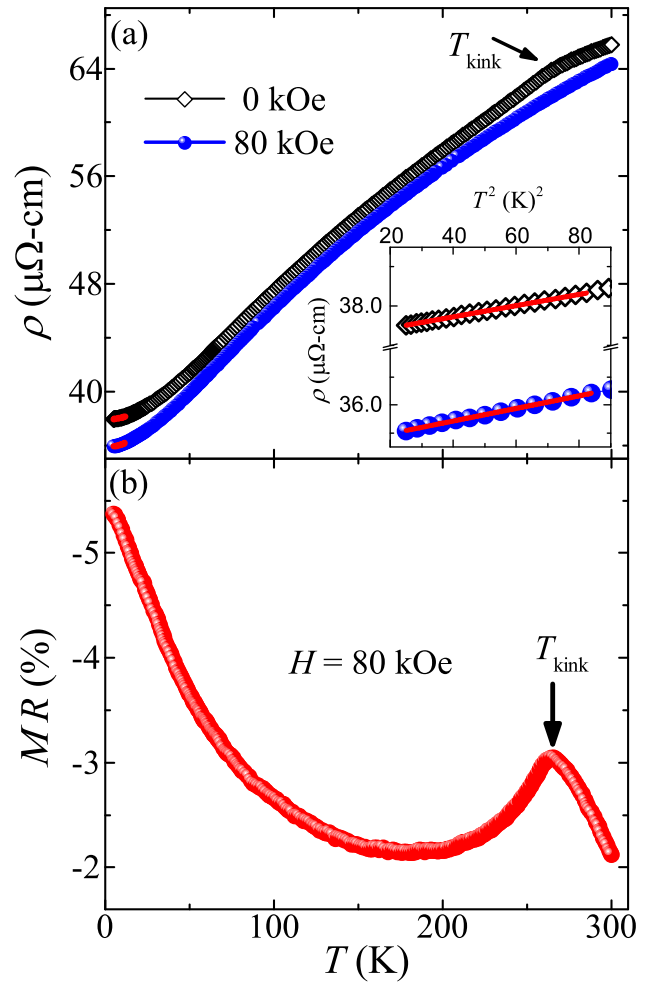}
	\end{center}
	\caption{\label{Fig7} (a) Temperature dependent resistivity [$\rho$(T)] in zero field and 80~kOe. Inset: $\rho$ vs $T^{2}$ at low temperatures highlighting the fit using Eq.~\eqref{eq.3}. (b) Magnetoresistance [$MR$(\%)] as a function of temperature in 80~kOe.}	     
\end{figure}
\begin{figure}[hbt]
	\begin{center}
		\includegraphics[width=\linewidth]{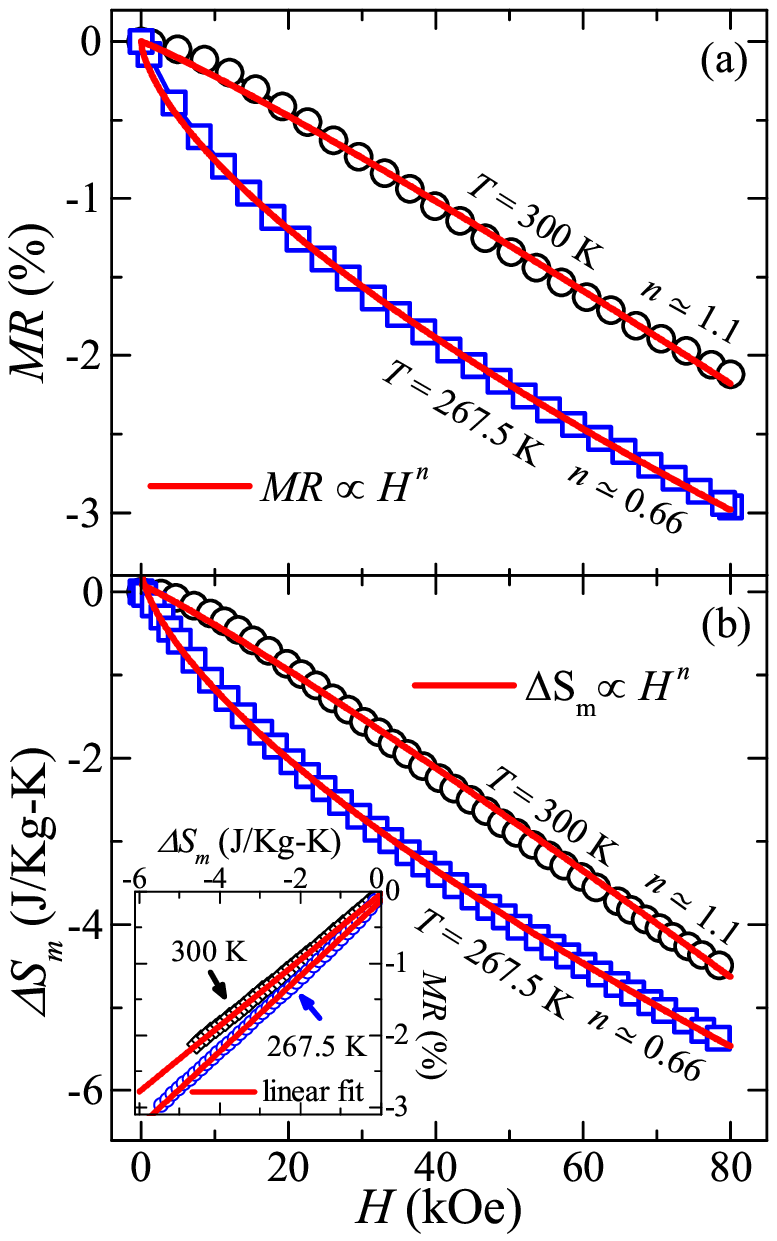}
	\end{center}
	\caption{\label{Fig8} (a) Magnetic field dependent $MR$(\%) at $T = 300$~K and 267.5~K. (b) $\Delta S_{\rm m}$ vs $H$ at $T = 300$~K and 267.5~K and the solid lines are the fits. Inset: Linear variation of $MR$(\%) with $\Delta S_{\rm m}$.}	     
\end{figure}

Typically, in a short-range model, the exchange interaction $J(r)$ decays rapidly with distance $r$ as $J(r)\sim e^{-r/\xi}$. On the other hand, for long-range interactions, the exchange interaction in $d$-dimension should decay following $J(r) \sim r^{-\rm (d+\sigma)}$ which is valid for the range of exchange interaction $(\sigma)<2$~\cite{Fisher917}. The final value of $\sigma = 1.41$ was obtained by examining the renormalization theory for the critical exponents choosing lattice dimensionality $d=2$ and spin dimensionality $n=1$ for $\gamma = 1.445$. The accuracy of these exponents are also tested by recalculating the other critical exponents (see the Supplementary Material). Thus, $d = 2$, $n = 1$, and $\sigma < 2$ reflect that the system belongs to a 2D Ising universality class with a long-range exchange interaction decaying as $J(r)\sim r^{-3.41}$. Our findings of the critical exponents are similar to that reported for ferromagnets Y$_{2}$Ni$_{7}$ and URhAl where the quasi-2D Ising character with long-range interactions is explained in term of strong anisotropy in the $ab$-plane and itinerant character of the magnetism~\cite{Bhattacharyya184414,Tateiwa064423}. Thus, the observed itinerant character of $3d$ electrons and strong anisotropy in the $ab$-plane at the transition temperature are the main sources of the quasi-2D Ising behavior with long-range interactions in Mn$_{1.4}$Fe$_{3.6}$Si$_{3}$ and other compounds in the Mn$_{1+x}$Fe$_{4-x}$Si$_{3}$ series.

\subsection{Resistivity and Magnetoresistance}
Figure~\ref{Fig7}(a) depicts the temperature dependent resistivity [$\rho(T)$] in zero magnetic field and 80~kOe in the temperature range 4~K to 310~K. The decrease of $\rho$ with $T$ reflects the metallic character of the compound. In zero field, $\rho(T)$ exhibits a weak anomaly or kink at $T_{\rm kink} \simeq 264.7$~K which corresponds to the PM-FM transition, similar to other metallic feromagnets~\cite{Samatham115118,Bhattacharyya184414,Yelland184436}. At low-temperatures, $\rho(T)$ is fitted by
\begin{equation}
\rho = \rho_{0} + A~T^{2}.
\label{eq.3}
\end{equation}
Here, $\rho_{0}$ is the residual resistivity and $A$ is the coefficient of $T^{2}$. The inset of Fig.~\ref{Fig7}(a) presents the linear variation of $\rho(T)$ with $T^{2}$ which is a representation of the Fermi-liquid behavior likely due to electron-electron scattering at low temperatures~\cite{Lohneysen1015,Stewart797}. Both the data sets below 9~K are fitted by Eq.~\eqref{eq.3} yielding ($\rho_{0} \simeq 37.91~\mu\Omega$~cm, $A \simeq 0.001434\mu\Omega$~cm/K$^{2}$) and ($\rho_{0} \simeq 35.852~\mu\Omega$~cm, $A \simeq 0.0022~\mu\Omega$~cm/K$^{2}$), in zero field and 80~kOe, respectively. The value of residual resistivity ratio is calculated to be RRR = $\rho_{300K}/\rho_{0} \simeq 1.83$ in zero field, indicating good quality of the sample.
Further, as the magnetic field is applied the residual resistivity $\rho_0$ is reduced significantly which could be related to the complex magnetic scatterings at low temperatures.

The temperature dependent magnetoresistance $\left[MR (\%)=\frac{\rho(H, T)- \rho(0, T)}{\rho (0, T)}\times 100\right]$ calculated from the resistivity data in zero field and in 80~kOe is presented in Fig.~\ref{Fig7}(b). The negative $MR$ is observed through out the temperature range (4-300~K) which increases with decreasing temperature, exhibits a broad hump around $T_{\rm kink}$ and then shows a gradual increase at low temperatures. A large and negative value of $MR \simeq -3\%$ at the transition temperature can be attributed to the suppression of spin disorder by the application of magnetic field~\cite{Hiroshi1828}. The isothermal $MR$ as a function of magnetic field measured near the transition ($T=267.5$~K) and at room temperature ($T = 300$~K) is shown in Fig.~\ref{Fig8}(a). The negative $MR$ increases with field and its value at $H = 80$~kOe matches with $MR(T)$ data in Fig.~\ref{Fig7}(b).

The isothermal $MR$ data are fitted by power law $MR(H)\propto H^n$. At $T = 300$~K, it exhibits almost a linear behavior with an exponent $n\simeq 1.1$, whereas a reduced exponent $n\simeq 0.66$ is obtained for $T = 267.5$~K. Interestingly, these values of $n$ are identical to the values obtained from the fit of $MCE$ data by $\Delta S_{m}(H) \propto H^{n}$, as shown in Fig.~\ref{Fig8}(b) and also consistent with the $n$ vs $T$ and $H$ plot in Fig.~S6 (Supplementary Material). Further, $MR$ as a function of $\Delta S_{m}$ with $H$ as an explicit parameter [inset of Fig.~\ref{Fig8}(b)] exhibits a linear behavior for both the temperatures. The identical field dependence of $MR$ and $MCE$ isotherms indicates that both the quantities are correlated and have same origin. This also suggests the itinerant character of $3d$ electrons in the studied compound. The identical behaviour of $MR$ and $MCE$ across the magnetic transition is previously reported in different intermetallic compounds with itinerant character~\cite{Singh046101,RawatL379,Rawat207,Campoy134410,Gupta012403,Samatham115118}.

\section{Conclusion}
The structural, magnetic, and electronic properties of Mn$_{1.4}$Fe$_{3.6}$Si$_{3}$ are investigated in detail by means of the temperature dependent powder XRD, magnetization, $MCE$, and $MR$ measurements. An anisotropic structural distortion is observed across the magnetic transition, where the lattice expands in the $ab$-plane and contracts slightly along the $c$-axis. This results in a large negative volume thermal expansion ($\alpha_{\rm V}\sim -20 \times 10^{-6}$~K$^{-1}$) across the transition temperature. The critical analysis of magnetization and $MCE$ data supports the second order nature of the phase transition which can be described in the framework of a quasi-2D Ising model with long-range magnetic interactions. The effective 2D character of the magnetic interactions originates from strong in-plane anisotropy across the transition while the itinerant or delocalized character of the $3d$ electrons is responsible for the long-range magnetic interactions in the Mn$_{1+x}$Fe$_{4-x}$Si$_{3}$ series.

\section{Supplementary Material} See the supplementary material for the critical analysis of magnetization and magnetocaloric effect data.

\section{Acknowledgments} We acknowledge BRNS, India for financial support bearing sanction Grant No.37(3)/14/26/2017. VS was supported by IISER Thiruvananthapuram postdoctoral program.

\section{Data Availability}
The data that support the findings of this study are available within the article [and its Supplementary Material].


\providecommand{\noopsort}[1]{}\providecommand{\singleletter}[1]{#1}%

\widetext
\clearpage
\begin{center}
	\textbf{\large Supplementary Material for\\}
	\textbf{\large "Negative thermal expansion and itinerant ferromagnetism in Mn$_{1.4}$Fe$_{3.6}$Si$_{3}$"}
\end{center}
\setcounter{equation}{0}
\setcounter{figure}{0}
\setcounter{table}{0}
\setcounter{page}{1}
\makeatletter
\setcounter{section}{0}
\renewcommand{\thesection}{S-\Roman{section}}
\renewcommand{\thetable}{S\arabic{table}}
\renewcommand{\theequation}{S\arabic{equation}}
\renewcommand{\thefigure}{S\arabic{figure}}
\renewcommand{\bibnumfmt}[1]{[S#1]}
\renewcommand{\citenumfont}[1]{S#1}
\title{Negative thermal expansion and itinerant ferromagnetism in Mn$_{1.4}$Fe$_{3.6}$Si$_{3}$} 
\author{Vikram Singh}
\email{vikram51128@gmail.com}
\affiliation{School of Physics, Indian Institute of Science Education and Research Thiruvananthapuram-695551, India}
\author{R. Nath}
\email{rnath@iisertvm.ac.in}
\affiliation{School of Physics, Indian Institute of Science Education and Research Thiruvananthapuram-695551, India}
\date{\today}



\section{Critical behaviour}
\subsection{Critical analysis of magnetization}

The critical behaviour analysis across a ferromagnetic transition temperature ($T_{\rm C}$) provides vital information about the phase transition and the exchange interaction. The critical behavior is characterized by a set of exponents ($\beta$, $\gamma$, and $\delta$) and the universality class of a magnetic system is typically decided on the basis of critical exponent values \textit{i.e.} mean-field model, Heisenberg model, Ising model, and 3D-XY model etc.\cite{Kaul5s}. These critical exponents are associated with spontaneous magnetization ($M_{\rm S}$), zero field inverse susceptibility ($\chi_{0}^{-1}$), and critical isotherm at $T_{\rm C}$ through the following expressions.\cite{Stanley336s}
\begin{equation}
	M_{\rm S}(T) = M_{0}(-\epsilon)^{\beta},~{\rm for}~\epsilon < 0, T<T_{\rm C},
	\label{eq.S1}
\end{equation}
\begin{equation}
	\chi_0 ^{-1}(T) = \Gamma(\epsilon)^{\gamma},~{\rm for}~\epsilon > 0,  T > T_{\rm C},
	\label{eq.S2}
\end{equation}
\centerline{and}
\begin{equation}
	M(H) = X(H)^{1/\delta},~{\rm for}~\epsilon = 0, T = T_{\rm C}.
	\label{eq.S3}
\end{equation}
Here, $\epsilon = \dfrac{T-T_{\rm C}}{T_{\rm C}}$ is the reduced temperature, $H$ is applied field, and $M_{0}$, $\Gamma$, and $X$ are the critical coefficients. These critical exponents are universal and related to each other through the following scaling laws \cite{Widom3898s}
\begin{equation}
	\alpha + 2\beta + \gamma = 2
	\label{eq.S4}
\end{equation}
\centerline{and}
\begin{equation}
	\delta = 1 + \frac{\gamma}{\beta}.
	\label{eq.S5}
\end{equation} 
Apart from these relations, the exponents should also satisfy the scaling equation of state which relates magnetization $M$ with $H$ and $T$ just above and below the $T_{\rm C}$ as
\begin{equation}
	M(H,\epsilon)\left\lvert\epsilon\right\rvert^{-\beta}=\textit{f}_{\pm}(H\left\lvert\epsilon\right\rvert^{-(\beta +\gamma)}).
	\label{eq.S6}
\end{equation}
Here, $\textit{f}_{+}$ and $\textit{f}_{-}$ are the scaling functions just above and below $T_{\rm C}$, respectively. The terms in the left and right hand sides of Eq.~\eqref{eq.S6} can also be written in terms of reduced magnetization $m = M(H,\epsilon)\epsilon^{-\beta}$ and reduced field $h = H\left\lvert\epsilon\right\rvert^{-(\beta +\gamma)}$, respectively. With the appropriate values of $\beta$, $\gamma$, and $T_{\rm C}$, the curves obtained from the implementation of Eq.~\eqref{eq.S6} will collapse into two separate universal branches: one above and another below $T_{\rm C}$.
\begin{figure*}[t]
	\begin{center}
		\includegraphics[width=15 cm]{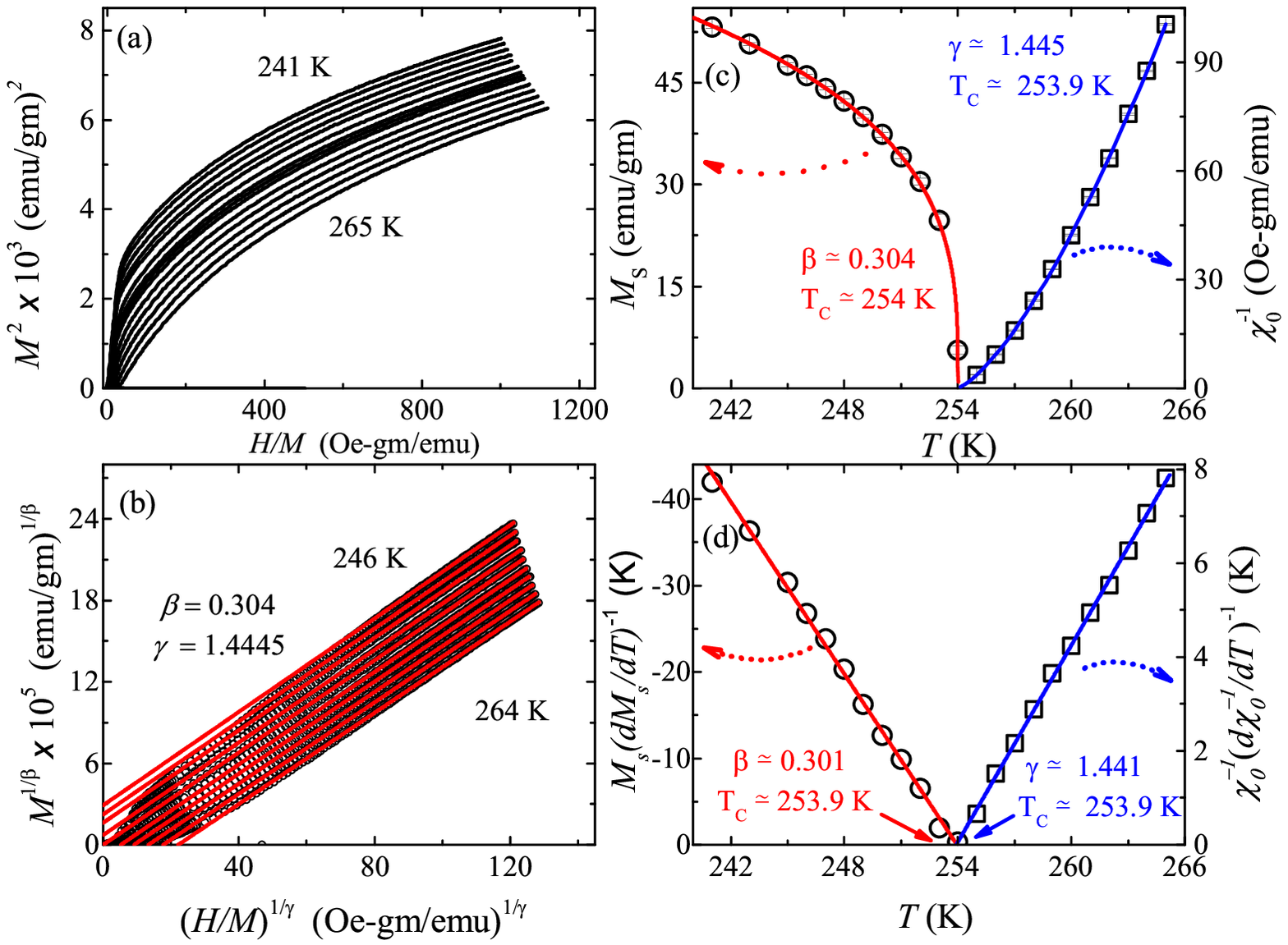}
	\end{center}
	\caption{\label{FigS1} (a) The Arrott plots: $M^{2}$ vs $H/M$. (b) The modified Arrott plots: $M^{1/\beta}$ vs $H/M^{1/\gamma}$. (c) Spontaneous magnetization ($M_{\rm S}$) and zero field inverse susceptibility ($\chi_{0} ^{-1}$) as a function of temperature in the left and right $y$-axes, respectively (d) The Kouvel-Fisher plot of $ M_{\rm S} $ and $ \chi_0^{-1}$ as a function of temperature. The solid lines are the linear fits. The solid arrows point to $T_{\rm C}$. }
\end{figure*}
\begin{figure}[t]
	\begin{center}
		\includegraphics[width=11 cm]{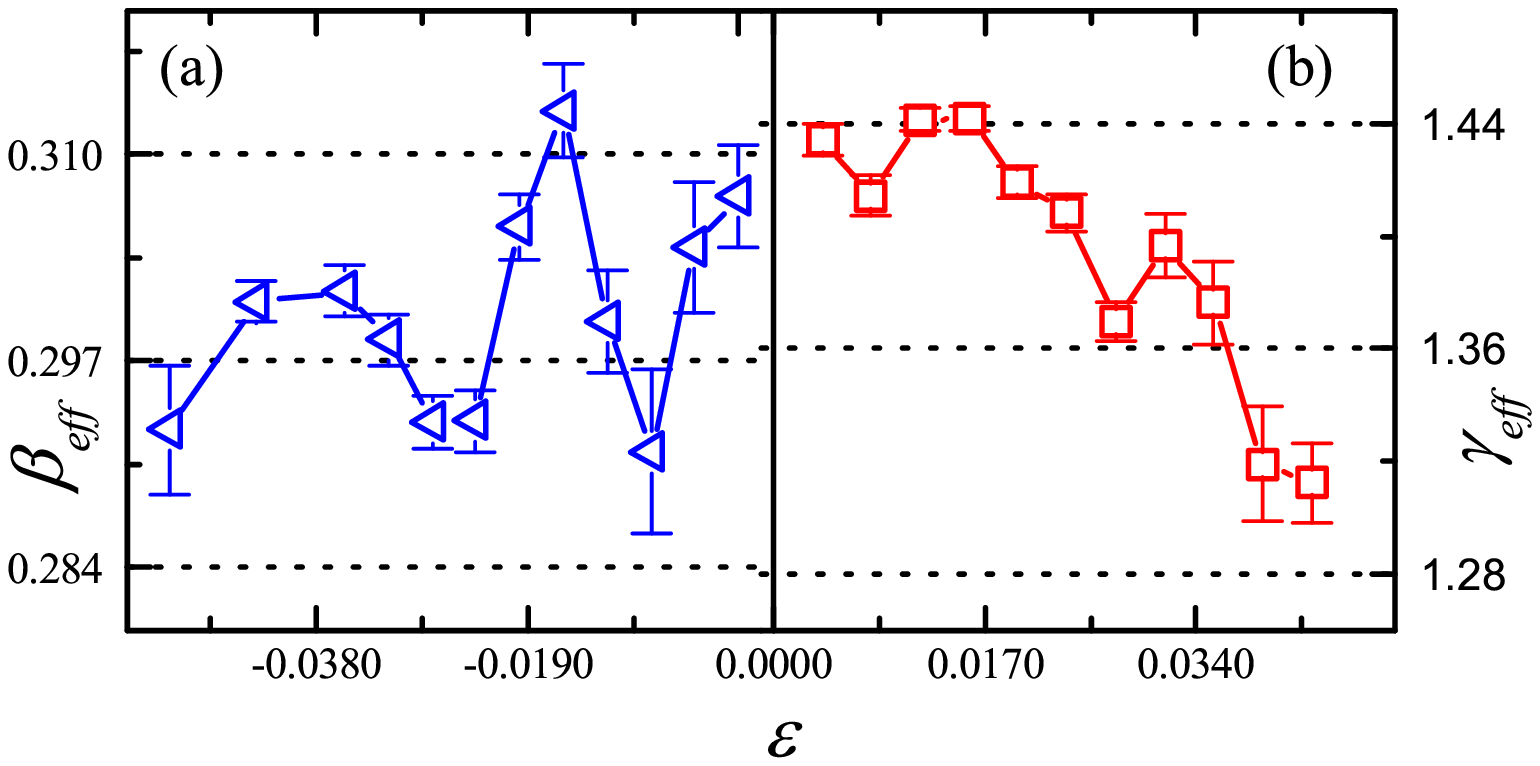}
	\end{center}
	\caption{\label{FigS2} The variation of effective exponents (a) $\beta_{\rm eff}$ and (b) $\gamma_{\rm eff}$ with reduced temperature ($\epsilon$).}
\end{figure}
\begin{figure}[hbt]
	\begin{center}
		\includegraphics[width=7 cm]{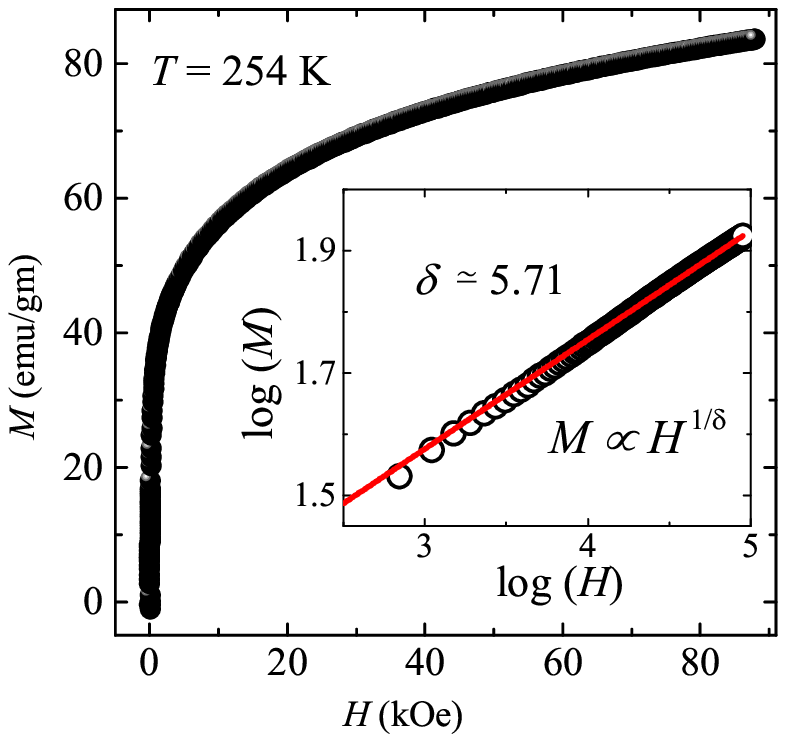}
	\end{center}
	\caption{\label{FigS3} Isothermal magnetization ($M$) vs applied field ($H$) at $T \simeq T_{\rm C} = 254$~K. Inset: Log$M$ vs log$H$ plot and the solid line is the linear fit corresponding to Eq.~\eqref{eq.S3}.}
\end{figure}
\begin{figure}[t]
	\begin{center}
		\includegraphics[width=13 cm]{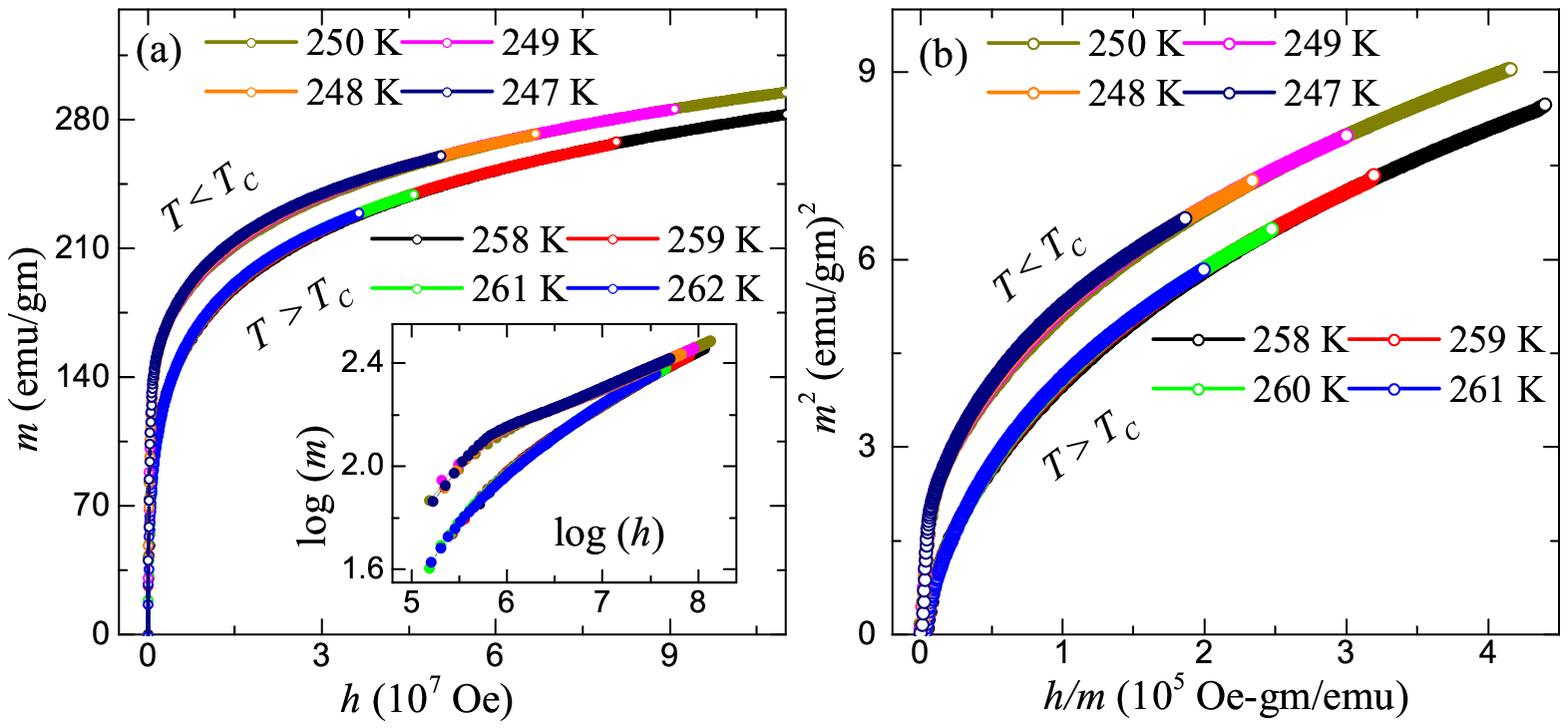}
	\end{center}
	\caption{\label{FigS4} (a) The reduced magnetization ($m$) vs reduced field ($h$) at different temperatures, just above and below $T_{\rm C}$. Inset: Log - log plots of the $m$ vs $h$ curves. (b) $m^2$ vs $h/m$ at the same temperatures around $T_{\rm C}$. The renormalized curves in (a) and (b) at different temperatures just above and below $T_{\rm C}$ are collapsing into two separate branches.}
\end{figure}
To determine the critical exponents from the magnetic isotherms we have employed the standard procedure. The conventional Arrott plots [$M^{2}$ vs $H/M$]\cite{Arrott1394s} are constructed from the magnetic isotherms $M(H)$ as depicted in Fig.~\ref{FigS1}(a). The positive slope of Arrott plots indicates a second order transition~\cite{Banerjee16s}. Clearly, the Arrott plots do not constitute parallel and straight lines as anticipated from the classical mean field model ($\beta=0.5$ and $\gamma=1.0$). Therefore, using the Arrott-Noakes equation, we employed modified Arrott plot (MAP) [$M^{1/\beta}$ vs $(H/M)^{1/\gamma}$] to further analyze the magnetic isotherms~\cite{Arrott786s}. For a reliable estimation of critical exponents, we have used the iterative procedure as described in Ref.~[\onlinecite{Pramanik214426s}]. As a starting point, the initial values of exponents are taken to be $\beta = 0.367$ and $\gamma = 1.388$ corresponding to the Heisenberg Model and iterations are performed till a stable set of critical exponents are achieved. The final values of critical exponents are obtained to be $\beta = 0.304$ and $\gamma = 1.4445$ which are used to construct the MAPs. The resultant straight and parallel lines are presented in Fig.~\ref{FigS1}(b). The linear fits of the final MAPs in the high field regime are extrapolated to $H/M = 0$ and the final temperature dependent saturation magnetization [$M_{\rm S}(T)$] and zero-field inverse susceptibility [$\chi_{0}^{-1}(T)$] are obtained from the intercepts in the $M^{1/\beta}$- and $(H/M)^{1/\gamma}$-axes, respectively. In Fig.~\ref{FigS1}(c), $M_{\rm S}(T)$ and $\chi_{0}^{-1}(T)$ are fitted by Eq.~\eqref{eq.S1} and Eq.~\eqref{eq.S2} which yield the parameters ($\beta \simeq 0.304$ and $T_{\rm C} \simeq 254$~K) and ($\gamma \simeq 1.445$ and $T_{\rm C} \simeq 253.9$~K), respectively.

To determine the more accurate values of parameters $\beta$, $\gamma$, and $T_{\rm C}$, we analyzed $M_{\rm S}(T)$ and $\chi_{0}^{-1}(T)$, obtained from the MAPs, using the Kouvel-Fisher (KF) method\cite{KouvelA1626s}:
\begin{equation}
	\dfrac{M_{S}(T)}{dM_{S}(T)/{dT}}=\dfrac{T-T_{C}}{\beta(T)}
	\label{eq.S7}
\end{equation}
\centerline{and}
\begin{equation}
	\frac{\chi_0^{-1}(T)}{d\chi_0^{-1}(T)/{dT}} = \dfrac{T-T_{C}}{\gamma(T)}.
	\label{eq.S8}
\end{equation}
Here, the exponents $\beta$ and $\gamma$ are temperature dependent and approach the exact values in the limit $T\rightarrow T_{\rm C}$. Following these equations, we plotted $M_{\rm S}(T)(dM_{\rm S}(T)/dT)^{-1}$ and $\chi_0 ^{-1}(T)(d\chi_0 ^{-1}(T)/dT)^{-1}$ vs temperature in Fig.~\ref{FigS1}(d) and the linear fits result $\beta\simeq 0.301$ with $T_{\rm C}\simeq 253.9$~K and $\gamma\simeq 1.441$ with $T_{\rm C}\simeq 253.9$~K, respectively. These values of $\beta$, $\gamma$, and $T_{\rm C}$ are consistent with the values determined from the MAP analysis.

Clearly, these critical exponents do not follow any conventional universality class.\cite{Singh6981s} Therefore, we analyzed the temperature variation of effective exponents ($\beta_{\rm eff}$ and $\gamma_{\rm eff}$) in order to check the presence of any crossover phenomena on approaching $T_{\rm C}$ which may gives rise to these unusual exponents. The effective exponents as a function of $\epsilon$ are obtained as:\cite{Pramanik214426s}
\begin{equation}
	\beta_{\rm eff}(\epsilon) = \dfrac{d[lnM_{S}(\epsilon)]}{{d(ln\epsilon)}},~\gamma_{\rm eff}(\epsilon) = \dfrac{d[ln\chi_{0}^{-1}(\epsilon)]}{{d(ln\epsilon)}}.
	\label{eq.S9}
\end{equation}
Figures~\ref{FigS2}(a) and (b) present $\beta_{\rm eff}$ and $\gamma_{\rm eff}$ as function of $\epsilon$, respectively. $\beta_{\rm eff}$ exhibits a nonmonotonic variation within 0.29 to 0.31 and seems to approach the observed $\beta \simeq 0.304$ value in the asymptotic limit ($\epsilon\rightarrow0$). Similarly, $\gamma_{\rm eff}$ increases slightly from 1.3 and converges to the observed value $\gamma \simeq 1.44$ for $\epsilon\rightarrow0$. These observations rule out the presence of any crossover regime near $T_{\rm C}$.

The third critical exponent ($\delta$) can be estimated from the Widom relation [Eq.~\eqref{eq.S5}] using the values of $\beta$ and $\gamma$, obtained from MAPs. Using $\beta \simeq 0.304$ and $\gamma \simeq 1.445$ in Eq.~\eqref{eq.S5}, we found $\delta \simeq 5.75$. $\delta$ can also be determined independently as per Eq.~\eqref{eq.S3} using the magnetic isotherm at $T_{\rm C}$. Figure~\ref{FigS3} contains the critical isotherm at $T_{\rm C} = 254$~K measured upto $H = 90$~kOe. The inverse slope of the linear fit to the log-log plot (see the inset of Fig.~\ref{FigS3}) of critical isotherm yields $\delta \simeq 5.71$, which is consistent with the value obtained from the Widom relation.

Finally, the obtained critical exponents should follow the scaling equation of state Eq.~\eqref{eq.S6} around $T_{\rm C}$. The plots of reduced magnetization $m$ as a function of reduced field $h$ using the actual exponents, should form two separate universal branches just above and below $T_{\rm C}$. The typical $m$ vs $h$ and $m^2$ vs $h/m$ plots are depicted in Fig.~\ref{FigS4}~(a) and (b), respectively. Indeed, the scaled magnetization isotherms of the studied compound follow Eq.~\eqref{eq.S6} and collapse into two separate branches just above and below $T_{\rm C}$. In the inset of Fig.~\ref{FigS4}(a), we have plotted $log~(m)$ vs $log~(h)$ to show that they exactly follow two branches even in the low field regime. Thus, the above observations unambiguously demonstrate the reliability of the obtained critical exponents and $T_{\rm C}$.

\subsection{Scaling and critical analysis of MCE}
\begin{figure}[hbt]
	\begin{center}
		\includegraphics[width=8 cm]{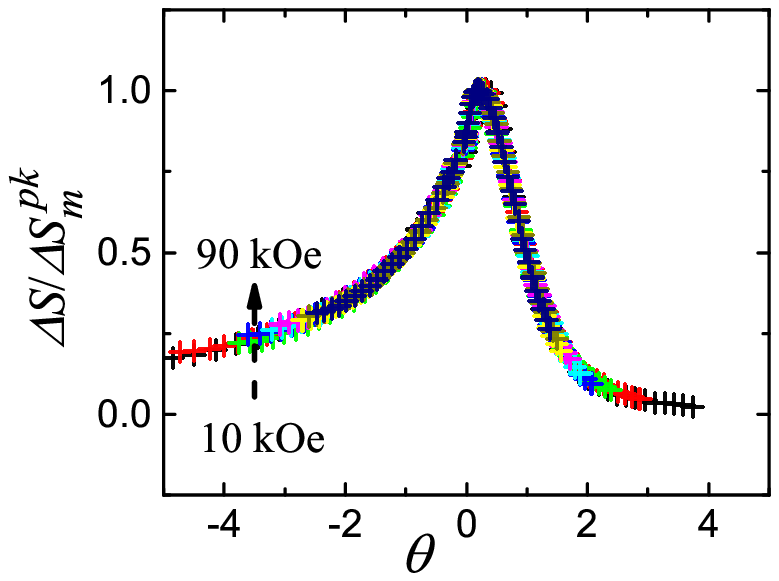}
	\end{center}
	\caption{\label{FigS5} Normalized magnetic entropy [$\Delta S_{\rm m}(T)/\Delta S_{\rm m}^{\rm pk}$] vs rescaled temperature $\theta$, obtained using Eq.~\eqref{eq.S10}.}
\end{figure}
Using the scaling hypothesis Franco~\textit{et~al}\cite{Franco222512s,Franco093903s} have shown that the $\Delta S_{\rm m}(T)$ curves at different magnetic fields should collapse on a single universal curve when the entropy change normalized to its peak value [$\Delta S_{\rm m}(T)/\Delta S_{\rm m}^{\rm pk}$] is plotted as a function of the rescaled temperature. This universality curve allows us to study the general behavior of $\Delta S_{\rm m}(T)$ for all applied fields. The rescaled temperature $\theta$ can be defined as
\begin{equation}
	\theta=\begin{cases}
		-(T-T_{\rm C})/(T_{\rm r1}-T_{\rm C}), & \text{for $T \le T_{\rm C}$}\\
		(T-T_{\rm C})/(T_{\rm r2}-T_{\rm C}), & \text{for $T > T_{\rm C}$},
	\end{cases}
	\label{eq.S10}
\end{equation}
where $T_{\rm r1}$ and $T_{\rm r2}$ are reference temperatures above and below $T_{\rm C}$ which should be chosen in such a way that the condition $\Delta S_{\rm m}$(for $T_{\rm r1} < T_{\rm C})/\Delta S_{\rm m}^{\rm pk} = \Delta S_{\rm m}$(for $T_{\rm r2} > T_{\rm C})/\Delta S_{\rm m}^{\rm pk} = h$ is satisfied. Here, $h$ is an arbitrary constant which has a value $h < 1$. To construct this phenomenological curve, we used $T_{\rm C}= 254$~K, obtained from the critical analysis of magnetization and reference temperatures ($T_{\rm r1}$ and $T_{\rm r2}$) corresponding to $h = 0.5$. As depicted in Fig.~\ref{FigS5}, a perfect collapse of all the curves is achieved in the full range of $\theta$. This finding is consistent with other compositions of the series Mn$_{1+x}$Fe$_{4-x}$Si$_{3}$ and reflects second order nature of the phase transition.\cite{Singh6981s} However, for a first order phase transition these curves do not collapse into a common curve, especially for $\theta < 0$.\cite{Bonilla224424s}

\begin{figure*}[hbt]
	\begin{center}
		\includegraphics[width=8 cm]{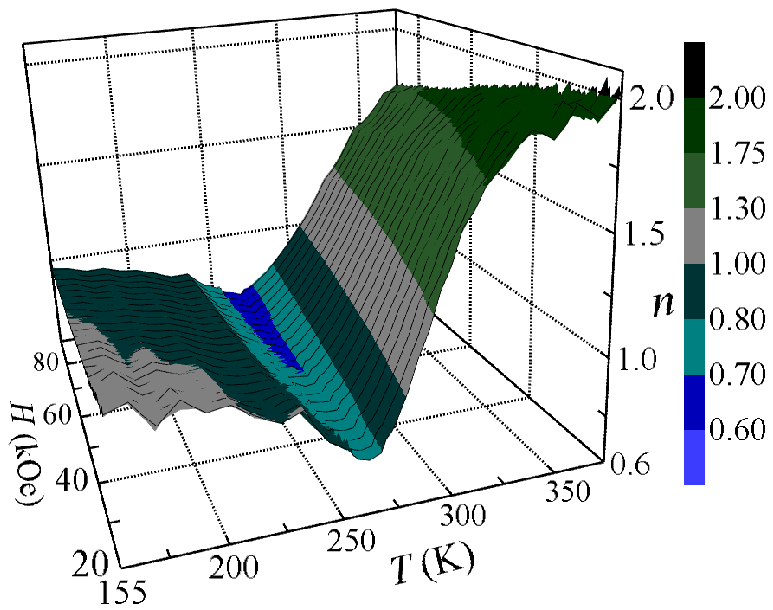}
	\end{center}
	\caption{\label{FigS6} Temperature and magnetic field dependent 3D plots of exponent $n$, obtained using Eq.~\eqref{eq.S12}.}
\end{figure*}
Furthermore, the critical exponents can be extracted from the power law fitting of the field dependent $MCE$ properties ($\Delta S_{\rm m}$, $\delta T_{\rm FWHM}$, and $RCP$) extracted from Fig.~6 (main text) in the vicinity of $T_{\rm C}$. It is demonstrated that magnetic field dependence of $\Delta S_{\rm m}$ can be expressed as: 
\begin{equation}
	\Delta S_{\rm m}~or~\rvert\Delta S_{\rm m}^{\rm pk}\rvert\propto H^{n}.
	\label{eq.S11}
\end{equation}
Here the exponent $n$ is temperature and field dependent which can be estimated locally from the following relation\cite{Shen5240s}:
\begin{equation}
	n(T, H) =\dfrac{d\ln|\Delta S_{\rm m}|}{d \ln H}.
	\label{eq.S12}
\end{equation}
In the case of a second order magnetic transition, this relation yields $n =2$ in the paramagnetic region ($T>>T_{\rm C}$) and $n = 1$ for $T<<T_{\rm C}$. However, in the critical/asymptotic regime (\textit{i.e.} $T = T_{\rm C}$ or $\epsilon \rightarrow 0$), $n$ is related to critical exponents $\beta$, $\gamma$, $\delta$, specific heat exponent ($\alpha$), and gap exponent ($\Delta$) as\cite{Franco285207s}
\begin{equation}
	n=1 + \frac{1}{\delta} \left(1-\frac{1}{\beta}\right)= 1+\dfrac{\beta-1}{\beta+\gamma}= (1-\alpha)/\Delta.
	\label{eq.S13}
\end{equation}

\begin{figure*}[hbt]
	\begin{center}
		\includegraphics[width= 13 cm]{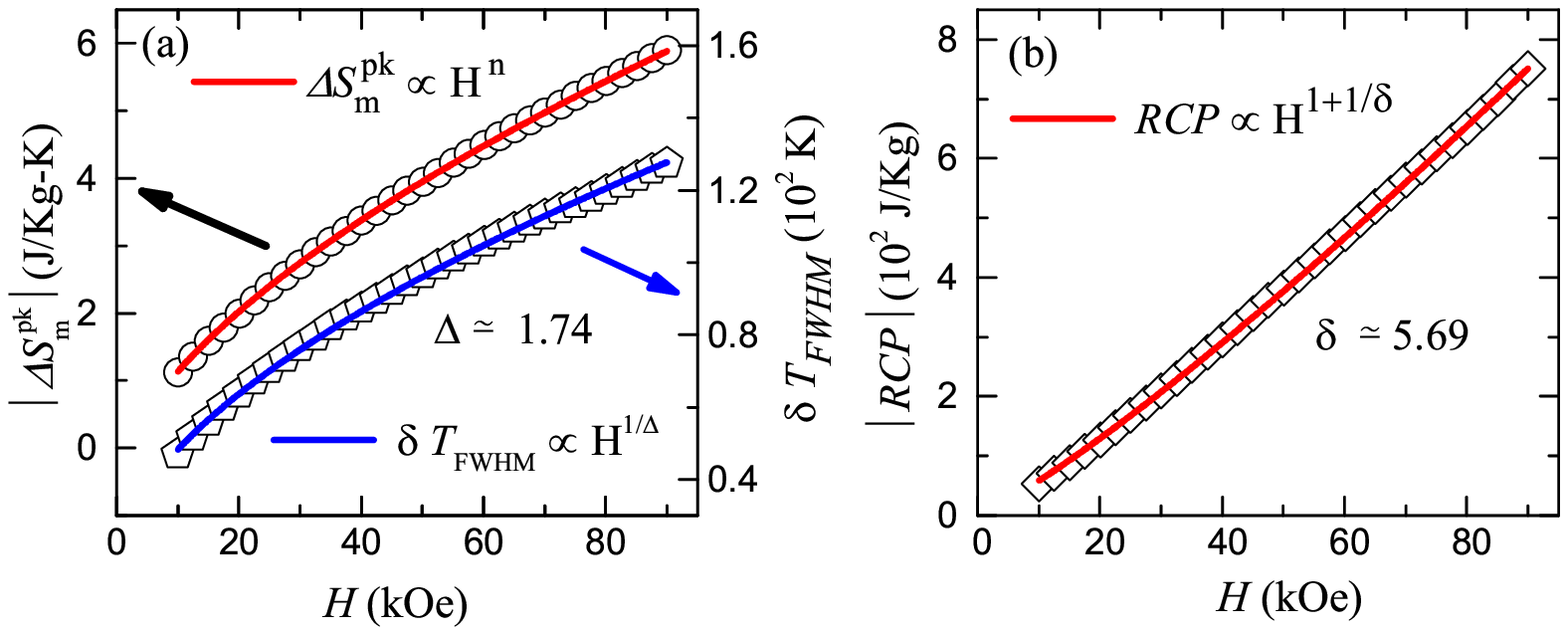}
	\end{center}
	\caption{\label{FigS7} The magnetic field dependent plot of (a) $\left| \Delta S_{\rm m}^{\rm pk}\right|$ (left $y$-axis) and $\delta T_{\rm FWHM}$ (right $y$-axis) and (b) relative cooling power ($RCP$) obtained from Fig.~6 (main text). The solid lines are the fits as described in the text.}
\end{figure*}

Figure~\ref{FigS6} presents the 3D plot of exponent $n$ as a function magnetic field and temperature calculated using of Eq.~\eqref{eq.S12}. It approaches value 2 and 1 in the paramagnetic region ($T>>T_{\rm C}$) and in the low temperature region ($T<<T_{\rm C}$), respectively. $n(T)$ exhibits a minima with $n \simeq 0.601$ at $T = T_{\rm C}$, typical behavior across a second order transition. Therefore, $n$ is an important exponent which provides information about the critical behavior in the asymptotic regime. Further, we have also estimated $n$ fitting $\Delta S_{\rm m}^{\rm pk} (H)$ by Eq.~\eqref{eq.S11} [see Fig.~\ref{FigS7}(a)] which yields $n \simeq 0.602$. This value of $n$ is consistent with the values obtained from Fig.~\ref{FigS6} at $T = T_{\rm C}$ and calculated using the $\beta$ and $\gamma$ values obtained from the critical analysis of magnetization in Eq.~\eqref{eq.S13}.

Similarly, the other exponents $\delta$ and $\Delta$ can be directly obtained from the fitting of magnetic field dependent $RCP$ and $\delta T_{\rm FWHM}$ using the following power laws\cite{Franco222512s,Franco093903s}
\begin{equation}
	\delta T_{\rm FWHM} \propto H^{1/\Delta}.
	\label{eq.S14}
\end{equation}
\centerline{and}
\begin{equation}
	RCP\propto H^{1 + 1/\delta}
	\label{eq.S15}
\end{equation}
The power law fitting of field dependent $\delta T_{\rm FWHM}$ and $RCP$ yield the $\Delta \simeq 1.74$ and $\delta\simeq 5.69$ as shown in Fig.~\ref{FigS7}. Indeed, the obtained gap exponent $\Delta \simeq 1.74$ is consistent with the relation $\Delta = \beta \times \delta = \beta +\gamma$. We have also estimated the specific heat exponent to be $\alpha \simeq -0.045$ using the $n$ and $\Delta$ values in Eq.~\eqref{eq.S13}.


The critical exponents obtained from the analysis of magnetization isotherms, $\Delta S_{\rm m}$, and $RCP$ are consistent with each other, reflecting the reliability of our analysis. The obtained critical exponents from different techniques are tabulated along with the theoretically expected ones in Table~I (main text). Clearly, our experimental values of critical exponents do not fall under any of the short-range model universality classes.\cite{Singh6981s} In a conventional short-range model, the exchange interaction $J(r)$ decays rapidly with distance $r$ following $J(r)\sim e^{-r/\xi}$. On the other hand, for long-range interactions, the exchange interaction in $d$-dimension should decay as $J(r) \sim r^{-\rm (d+\sigma)}$ which is valid for the range of exchange interaction $(\sigma)<2$.\cite{Fisher917s} According to renormalization theory, critical exponent $\gamma$ can be defined as\cite{Fisher917s}
\begin{equation}
	\begin{split}
		\gamma = 1 +\dfrac{4}{d}\Big(\dfrac{n+2}{n+8}\Big)\Delta\sigma + \dfrac{8(n+2)(n-4)}{d^{2}(n+2)^{2}}\\
		=\times\Big[1 + \dfrac{2G(\frac{d}{2})(7n+20))}{(n-4)(n+8)}\Big]\Delta\sigma^{2},
	\end{split}
	\label{eq.4}
\end{equation}
where $\Delta \sigma = (\sigma-\frac{d}{2})$, $ G(\frac{d}{2}) = 3-\frac{1}{4}(\frac{d}{2})^{2}$. Using the experimental value of $\gamma$ for different sets of ($d$, $n$) in Eq.~\eqref{eq.4} we estimated the corresponding $\sigma$ values. After various trails with different sets of ($d$, $n$) values, a reliable value of $\sigma = 1.41$ was obtained choosing $d=2$ and $n=1$ for $\gamma = 1.445$. In order to test the reliability of the $\sigma$ value, one can recalculate the other critical exponents as $\nu = \gamma/\sigma$, $\eta = 2-\sigma$, $\alpha = 2-\nu d$, $\beta = (2-\alpha-\gamma)/2$, and $\delta = 1 + \gamma/\beta$.\cite{Fisher917s,Fischer064443s} Using $\sigma = 1.41$ in the above relations we estimated $\beta \simeq 0.300$, $\gamma \simeq 1.448$, $\delta \simeq 5.831$, $\nu \simeq 1.02$, $\eta \simeq0.586$, and $\alpha \simeq-0.0475$ which are consistent with our experimental values, justifying our choice of ($d$, $n$) and the estimation of $\sigma = 1.41$. Thus, with $d = 2$, $n = 1$, and $\sigma < 2$ the system belongs to a 2D Ising universality class with a long-range exchange interaction decaying as $J(r)\sim r^{-3.41}$.

\end{document}